\documentstyle[12pt]{article}

\def\MeV{\,{\rm MeV}}
\def\sec{\,{\rm sec}}

\def\Mpc{\,{\rm Mpc}}

\def\cmm2{{\,\rm cm^{-2}}}
\def\cm2{{\,{\rm cm}^2}}
\def\cmm3{{\,{\rm cm}^{-3}}}
\def\gcmm3{{\,{\rm g\,cm^{-3}}}}
\def\kms{\,{\rm km\,s^{-1}}}

\def\la{\mathrel{\mathpalette\fun <}}
\def\ga{\mathrel{\mathpalette\fun >}}
\def\fun#1#2{\lower3.6pt\vbox{\baselineskip0pt\lineskip.9pt
  \ialign{$\mathsurround=0pt#1\hfil##\hfil$\crcr#2\crcr\sim\crcr}}}

\begin{document}
\pagestyle{empty}
\begin{center}
\rightline{FERMILAB--Pub--94/059-A}
\rightline{astro-ph/9403054}
\rightline{submitted to {\it Nature}}

\vspace{.1in}
{\Large \bf RELAXING THE BIG BANG BOUND \\
\bigskip
TO THE BARYON DENSITY} \\

\vspace{.2in}
Geza Gyuk$^{1,2}$ and Michael S. Turner$^{1,2,3}$\\
\vspace{.2in}

{\it $^1$Department of Physics\\
The University of Chicago, Chicago, IL~~60637-1433}\\

\vspace{.1in}

{\it $^2$NASA/Fermilab Astrophysics Center\\
Fermi National Accelerator Laboratory, Batavia, IL~~60510-0500}\\

\vspace{0.1in}

{\it $^3$Department of Astronomy \& Astrophysics\\
Enrico Fermi Institute, The University of Chicago,
Chicago, IL~~60637-1433}\\

\end{center}

{\bf In the standard picture of big-bang nucleosynthesis the
yields of D, $^3$He, $^4$He, and $^7$Li only agree with
their inferred primordial abundances if the fraction of
critical density contributed by baryons is between $0.01h^{-2}$
and $0.02h^{-2}$ ($h$ is the present value of the Hubble constant
in units of $100\kms\Mpc^{-1}$).  This is the basis of the
very convincing and important argument that baryons can contribute
at most 10\% of critical density and thus cannot close the Universe.
Nonstandard scenarios involving decaying particles,$^1$ inhomogeneities
in the baryon density,$^2$ and even more exotic ideas$^3$
put forth to evade this bound have been largely unsuccessful.$^4$
We suggest a new way of relaxing the bound:   If the tau neutrino has a
mass of $20\MeV-30\MeV$ and lifetime of $200\sec -1000\sec$, and its
decay products include electron neutrinos, the bound to the
baryon mass density can be loosened by a about factor of $10$.
The key is the decay-generated electron antineutrinos:
around the time of nucleosynthesis they are captured by protons to
produce neutrons, thereby changing the outcome of nucleosynthesis.
Experiments at $e^\pm$ colliders should soon be sensitive to a
tau-neutrino mass in the required range.}

\newpage
\pagestyle{plain}
\setcounter{page}{1}

Big-bang nucleosynthesis is one of the cornerstones of
the hot big-bang cosmology.  The successful prediction
of the primordial abundances of D, $^3$He, $^4$He, and
$^7$Li tests the standard cosmology back to the epoch
of nucleosynthesis ($t\sim 0.01\sec - 300\sec$ and
$T\sim 1\MeV -0.03\MeV$),
providing its earliest check (see Refs.~5).  It also poses an important
challenge:  the identification of the
ubiquitous dark matter that is known to account for most
of the mass density and plays a crucial role in the formation
of structure in the Universe.

Standard big-bang nucleosynthesis
implies that the fraction of critical density
contributed by baryons must be
between $0.01h^{-2}$ and $0.02h^{-2}$, or
less than about 10\% for a generous range in the
Hubble constant.$^6$  Taken together with the fact
that luminous matter contributes much less than 1\% of the
critical density this leads to the following possibilities:
(i) $\Omega_B=\Omega_0 \la
0.1$, the dark matter is baryonic and the Universe is open;
(ii) $\Omega_0 \ga 0.1\ga \Omega_B$ and
much, if not most, of the dark matter is something other than
baryons; or (iii) the standard picture of nucleosynthesis is
somehow wrong or incomplete, and $\Omega_B$ is greater
than 0.1.  ($\Omega_0$ is the ratio of the total mass density
to the critical density.)

The first possibility, an open, baryon-dominated Universe,
is certainly not precluded.  However, a number of lines
of reasoning suggest that $\Omega_0$ is larger than 0.1,
perhaps even as large as unity.  Measurements of $\Omega_0$
based upon cluster dynamics, based upon the ratio of total mass
to baryonic mass in clusters, and those based upon relating the peculiar
motions of galaxies to the observed distribution of matter all
strongly favor a value for $\Omega_0$ that is much larger than 0.1.$^7$
Further, the most successful models for structure formation are predicated
upon a flat Universe whose dominant form of matter is slowly
moving elementary particles left over from the earliest
moments (``cold dark matter'').$^8$
Even in the one viable model of structure formation
with baryons only (PIB$^9$), the nucleosynthesis bound is violated
by a large margin, $\Omega_B = \Omega_0 \sim 0.2-0.4$ with $h\sim 0.5-0.8$.
Finally, theoretical prejudice, most especially the
Dicke-Peebles timing argument$^{10}$ and inflation,$^{11}$ strongly favor
a flat Universe.

If $\Omega_0$ does exceed 0.1, one is
pushed either to option (ii), a new form of matter, or to option (iii),
a modification of the standard picture of primordial nucleosynthesis.
While previous attempts to circumvent the
nucleosynthesis bound to $\Omega_B$ have been unsuccessful,$^4$
the alternative for $\Omega_0\ga 0.1$ is a radical one.
Thus, we believe that it is worth exploring
modifications to the standard picture, especially when they
are testable, as is the one discussed here.

To begin, let us review what goes wrong with the light-element abundances
for large $\Omega_B$.  The yields of nucleosynthesis
depend upon the baryon-to-photon ratio $\eta$ which is related
to $\Omega_B$ by
\begin{equation}
{\Omega_Bh^2\over 0.1} = {\eta \over 27\times 10^{-10}}.
\end{equation}
For the standard picture of nucleosynthesis the concordance
range is $\eta \simeq 3-5\times 10^{-10}$.
At the time of nucleosynthesis (around
$200\sec$) essentially all neutrons are incorporated into
$^4$He.  However, because of decreasing particle densities and
decreasing temperature nuclear reactions eventually
cease (``freeze out'') and a small fraction of
the neutrons remain in D and $^3$He.  Nuclear reaction rates
(per particle) depend directly upon $\eta$; for this reason
the $^4$He mass fraction increases with $\eta$, though only
logarithmically since the additional $^4$He synthesized is
small.  The D and $^3$He yields depend more sensitively upon $\eta$,
decreasing as a power of $\eta$.

The $^7$Li story is more complicated; the key to understanding
it involves the free-neutron fraction around the time of
nucleosynthesis.  When it is relatively large, as for $\eta \la
3\times 10^{-10}$, $^7$Li is produced by $^4$He(t,$\gamma$)$^7$Li
and destroyed by $^7$Li(p,$\alpha$)$^4$He.  The final
$^7$Li abundance is determined by a competition between production
and destruction and decreases with increasing $\eta$
as production decreases (fewer neutrons and
less t) and destruction increases (larger $\eta$ results
in faster rates).  When the neutron fraction is relatively
low, as for $\eta\ga 3\times 10^{-10}$, $^7$Li is produced as
$^7$Be which, long after nucleosynthesis, $\beta$-decays to $^7$Li
via electron capture.  In this regime, the production
process is $^3$He($^4$He,$\gamma$)$^7$Be, and the destruction
process is $^7$Be(n,p)$^7$Li followed by $^7$Li(p,$\alpha$)$^4$He.
The yield of $^7$Be increases with increasing $\eta$ because
the production rate increases (larger $\eta$ leads to faster
rates) and the destruction rate decreases (fewer neutrons).
In the intermediate regime, $\eta \sim 3\times 10^{-10}$,
$^7$Li production achieves its minimum ($^7$Li/H $\sim 10^{-10}$)
and both $^7$Li and $^7$Be processes are important.

The problem with large $\eta$ is the overproduction of $^4$He
and $^7$Li and the underproduction of D.
To be more specific, if we take 0.25 as an upper bound to the primordial
mass fraction of $^4$He ($\equiv Y_P$), then
$\eta$ must be less than
$10\times 10^{-10}$.  Since $^4$He production increases
very slowly with $\eta$, taking instead $Y_P\le 0.255$ relaxes
the upper bound to $\eta$ significantly, to $20\times10^{-10}$.
Further, if one were to suppose
that the tau neutrino were massive ($m_\nu \gg 1\MeV$) and
disappeared before the epoch of nucleosynthesis so that the
number of light neutrino species was effectively two,
then $^4$He production constrains
$\eta$ to be less than $50\times 10^{-10}$ for $Y_P\le 0.25$,
and less than $75\times 10^{-10}$ for $Y_P\le 0.255$.

Deuterium is a much more sensitive ``baryometer.''  Since
there is no plausible
astrophysical source for D, big-bang production must account
for at least what is observed, D/H $\ga 10^{-5}$.
This results in the upper bound to $\eta$ of
$8\times 10^{-10}$.  Because D production
decreases so rapidly with $\eta$, this upper bound to $\eta$
is relatively insensitive to the assumed lower bound for D/H.

Finally, it is believed that the $^7$Li
abundance measured in the pop II halo stars, $^7$Li/H$\simeq
1.2\pm 0.3\times 10^{-10}$, accurately reflects the
primordial $^7$Li abundance.$^{6,12}$
Insisting that the $^7$Li yield be no greater than $^7$Li/H$= 1.5\times
10^{-10}$ implies an upper bound to $\eta$ of $4\times 10^{-10}$.
We note that there are still uncertainties in key
reaction rates for $^7$Li and in the interpretation of
the astrophysical measurements (has $^7$Li been astrated?;
different stellar atmosphere models lead to different $^7$Li
abundances for the same line strengths; and so on).$^{6,12}$
Even so, $^7$Li still poses a serious constraint:
taking instead $^7$Li/H$\la 10^{-9}$
only loosens the bound to $\eta \le 9\times 10^{-10}$.

In sum, the toughest challenge in relaxing
the big-bang bound is simultaneously addressing the
underproduction of D and the overproduction of $^7$Li.
As we now describe $10\MeV -30\MeV$ tau neutrino which
decays to electron neutrinos and
has a lifetime of order $300\sec$ can do just that!

Let us briefly review our recent detailed study of the effects of a
massive, unstable tau neutrino on primordial nucleosynthesis.$^{13}$
The abundance of a $20\MeV -30\MeV$ tau neutrino (per comoving volume)
ceases to decrease and freezes out when the temperature of
the Universe is a few MeV; until tau neutrinos decay, their abundance
per comoving volume remains constant.  For this mass
range the freeze-out abundance (assuming the
annihilation rate predicted in the standard electroweak model) is given by
$rm_\nu \sim 0.6\MeV - 1\MeV$,
where $r$ is the abundance relative to a massless neutrino species.
The quantity $rm_\nu$ serves to quantify the energy density; until
tau neutrinos decay their energy density,
$\rho_\tau (T) = rm_\nu n_\nu \simeq ({rm_\nu /3T}) \rho_{\nu 0}$,
where $\rho_{\nu 0}$ is the energy density of a massless neutrino species.

A decaying tau neutrino can have several effects on nucleosynthesis:$^{13}$
(i) the energy density it and its daughter products contribute
speed up the expansion
rate, tending to increase $^4$He production; (ii) if its decay
products include particles that interact electromagnetically
its decays increase the entropy density and thereby reduce
the baryon-to-photon ratio, which leads to decreased
$^4$He production and increased D production;
(iii) if its decay products include
electron neutrinos (and antineutrinos) their interactions with
nucleons affect the neutron-to-proton ratio and thereby
the outcome of nucleosynthesis.$^{14}$  In general, when the effects
of a decaying tau neutrino are significant they are deleterious
and large regions of the mass-lifetime plane can be
excluded on this basis.$^{15}$  There are exceptions; elsewhere$^{16}$
we discussed the potential beneficial effects of a $1\MeV -10\MeV$
tau neutrino for the cold dark matter scenario of structure
formation; here, we discuss another.

The decay modes of interest involve electron
neutrinos (and antineutrinos);
e.g., $\nu_\tau \rightarrow \nu_e \phi$ or $\nu_e +\nu_e{\bar\nu}_e$
(where $\phi$ is a very light pseudoscalar particle).
For the masses, $20\MeV -30\MeV$, lifetimes, $\tau_\nu \ga 200\sec$,
and the abundances of interest, the energy density
contributed by a massive tau neutrino is much less than that of
a massless one (effectively, the number of massless
neutrino species is two).  The main difference then, between standard
nucleosynthesis and that with a decaying tau neutrino, are
the electron neutrinos and antineutrinos that
are produced in equal numbers by tau-neutrino decays.
Because their energies are
much greater than the neutron-proton mass difference, $E_\nu
= (0.33\ {\rm or}\ 0.5)m_\nu \ga 6\MeV$, the capture cross section for
an antineutrino on a proton, ${\bar\nu}_e + p \rightarrow e^+ + n$,
is essentially equal to that for a neutrino on a neutron,
$\nu_e + n \rightarrow e^- + p$.
However, after the freeze out of the neutron-to-proton ratio,
which occurs when $T\sim 1\MeV$ and $t\sim 1\sec$,
protons outnumber neutrons
by about six to one, and so the capture of decay-produced neutrinos
will produce about six times as many neutrons as protons.

The probability that a nucleon captures an electron
neutrino or antineutrino around the time of nucleosynthesis
($T\sim 0.1\MeV$ and $t\sim 200\sec$) is proportional
to (capture cross section) $\times$ (number density of tau neutrinos)
$\times$ ($t\sim 200\sec$); more precisely,
\begin{equation} \label{eq:probability}
{\cal P} \approx 10^{-4} \left({rm_\nu \over 0.05\MeV}\right)
\left( {m_\nu \over 20\MeV}\right)\left( {200\sec \over \tau_\nu}\right) ,
\end{equation}
where we assume that $\tau_\nu \ga 200\sec$ so that the fraction
of tau neutrinos
that decay in a Hubble time around $t\sim 200\sec$ is $200\sec
/\tau_\nu$.  The upshot is that
tau-neutrino decays continuously produce neutrons
around the time of nucleosynthesis, amounting to a total of about
$10^{-4}$ per baryon.  This is the key to obtaining the ``correct''
D and $^7$Li abundances for large $\eta$:  the
neutron fraction is increased to the value that it would
have for much smaller $\eta$ (see Fig.~1), whereas in the standard
picture, for large $\eta$ neutrons are very inefficiently
incorporated into $^4$He, resulting in little D production
and few free neutrons, which leads to the overproduction of $^7$Li.

Because a decaying tau neutrino leads to a neutron fraction
that is very similar to that in the standard picture
with $\eta \sim 3\times 10^{-10}$, the yields of D and
$^7$Li are very similar and vary
only slowly with $\eta$ (see Fig.~2).
In the end, the maximum value of $\eta$ consistent with the
light-element abundances is controlled by the overproduction
of $^4$He (see Fig.~3).  Increasing $rm_\nu$ allows acceptable D and
$^7$Li abundances for larger and larger values of $\eta$;
however, it also increases the energy density contributed
by the tau neutrino and its decay products which increases
$^4$He production (from much less than that of a massless
neutrino species for $rm_\nu \la 0.1\MeV$ to close
to that of a massless neutrino species for $rm_\nu \sim 0.5\MeV$).
For tau neutrino masses between $10\MeV$ and $30\MeV$ the highest values of
$\eta$ consistent with the light-element abundances
are around $50\times 10^{-10}$ and occur for $rm_\nu\sim
0.03\MeV -0.1\MeV$ (see Fig.~4).  (Our criteria for
concordance are:  $Y_P \le 0.25$,
D/H$\ge 10^{-5}$, D+$^3$He/H$\le 10^{-4}$, and $0.5\times 10^{-10}
\le ^7$Li/H$\le 2\times 10^{-10}$.  For tau-neutrino
lifetimes $3000\sec \ga \tau_\nu \ga 200\sec$
our results are relatively insensitive to $\tau_\nu$ and
depend only slightly on decay mode.)

The maximum value of the baryon density that can be allowed
with a decaying tau neutrino is $\Omega_Bh^2 \simeq 0.2$; this
permits closure density in baryons for a Hubble constant of
slightly less than $50\kms\Mpc^{-1}$.  This absolute
bound to $\Omega_Bh^2$ rises to about $0.3$ when the
constraint to the primordial mass fraction of $^4$He is
relaxed to $Y_P\le 0.255$.

Loosening the nucleosynthesis bound to the baryon density
has manifold implications,
especially for the formation of structure in the Universe.
It makes the PIB model$^{9}$ consistent with the
nucleosynthesis bound to the baryon density.
Or, it allows a critical Universe with no exotic dark
matter.  If, in addition, much of the baryon mass formed into
massive objects early on, as suggested by some,$^{17}$ such a
scenario would have all the virtues of cold dark matter without
the necessity of a new form of matter.
Finally, increasing the baryon fraction to 20\% or 30\% of critical,
but maintaining the bulk of the mass density in cold dark
matter, leads to a version of ``mixed dark matter''
discussed a few years ago,$^{18}$ which has
the benefit of additional power on large scales.

Of course, increasing the number of baryons in the Universe
also raises some serious questions.  For example, where are all the baryons?
Recall, luminous matter contributes less than 1\% of the
critical density.  While the Gunn-Petersen test tells us that
essentially 100\% of the baryons in the IGM must be ionized,
the stringent COBE FIRAS limit to the Compton $y$ parameter
tell us that that there cannot be too much hot gas.$^{19}$

How plausible is a tau neutrino of mass $20\MeV$ to $30\MeV$
with a lifetime of order a few hundred seconds whose decay
products include electron neutrinos?  There is almost universal
belief that neutrinos have mass---and almost as many neutrino
mass schemes as there are particle theorists.  Such a decay mode and
lifetime can arise in models with ``family symmetries'' that
relate the quarks and leptons of different generations, or
models with additional $Z$ bosons.  In models where tau-neutrino
decays respect an $SU(2)$ symmetry, the decay width for the
charged tau-lepton decay mode $\tau \rightarrow 3e$
is related directly to that for the
tau-neutrino decay mode $\nu_\tau \rightarrow 3\nu_e$,
\begin{equation}
\Gamma (\tau \rightarrow 3e) = \left( {m_\tau \over m_\nu}\right)^5
\Gamma (\nu_\tau \rightarrow 3\nu_e) \simeq 6\times 10^6\sec^{-1}
\,\left({m_\nu \over 25\MeV}\right)^{5}
\left( {\tau_\nu\over 300\sec}\right)^{-1}.
\end{equation}
The current upper limit to the decay width for this mode
is $\Gamma (\tau\rightarrow 3e) \le
10^8\sec^{-1}$; improving the sensitivity by a factor of
10 or so offers a possible test of the $3\nu_e$ decay mode.

A careful reader will have noticed
that the value of the relic abundance needed to loosen the
nucleosynthesis bound, $rm_\nu \sim 0.03\MeV -0.1\MeV$, is about a factor
of ten smaller than that which results if the tau-neutrino
abundance is determined by the freeze out of annihilations
as predicted in the standard electroweak model.  However,
if neutrinos have mass, they necessarily have new
interactions (neutrino masses
are forbidden in the standard electroweak model).
Additional interactions increase the annihilation
cross section (a factor of about ten is required),
which decreases the tau-neutrino abundance.
Alternatively, entropy production after
the freeze out of the tau neutrino's abundance could have
reduced its abundance.  We hesitate to mention the
obvious generalization of our results:  if the tau neutrino
can't do the job, another particle with similar, or even greater mass,
and relic abundance $rm\sim 0.3\MeV -0.1\MeV$, could.

Finally, what are the prospects for testing this scenario?
Current laboratory upper
limits to the mass of the tau neutrino, based upon end-point
studies of tau decays to final states with five pions,
are just above $30\MeV$.$^{20}$
Prospects for improving the sensitivity of these experiments,
which are done at $e^\pm$ colliders, are good.
Thus, the uncertainty in the nucleosynthesis bound to the
baryon density raised here should be clarified in the near future.

\bigskip\bigskip\bigskip
We are very pleased to thank Scott Dodelson for many
helpful and enlightening conversations.
This work was supported in part by the DOE
(at Chicago and Fermilab) and by the NASA
through grant NAGW-2381 (at Fermilab).
GG is supported in part by an NSF predoctoral fellowship.

\vfill\eject
\section*{Figure Captions}
\bigskip

\noindent{\bf Figure 1:}  The neutron fraction as a function
of temperature in the standard scenario with $\eta =
3\times 10^{-10}$ (broken curve) and $\eta = 5\times 10^{-9}$
(dotted curve), and in the decaying tau-neutrino
scenario with $\eta =5\times 10^{-9}$, $m_\nu = 30\MeV$,
$\tau_\nu =300\sec$, and $rm_\nu = 0.03\MeV$ (solid curve).

\medskip
\noindent{\bf Figure 2:}  The abundance of D and $^7$Li as a
function of $\eta$ in the standard scenario (solid curves)
and with a decaying
tau neutrino (broken curves; $rm_\nu = 0.03\MeV$, $m_\nu =
30\MeV$, and $\tau_\nu =400\sec$).

\medskip
\noindent{\bf Figure 3:}  The maximum value of $\eta$ consistent
with D, $^7$Li, and $^4$He (solid:  $Y_P\le 0.25$;
broken:  $Y_P\le 0.255$) production shown separately
as a function of $rm_\nu$ for a tau neutrino of mass $20\MeV$
and lifetime $\tau_\nu = 300\sec$.  The curve labeled D+$^3$He
and the lower $^7$Li curve are lower limits to $\eta$.

\medskip
\noindent{\bf Figure 4:}  The maximum value of $\eta$ consistent
with all the light-element abundances as a function of $rm_\nu$
for $m_\nu= 10, 15, 20, 25 \MeV$ and $\tau_\nu =300\sec$.
The peak of the curve moves from right to left with increasing mass.

\end{document}